\documentclass{aa}
\usepackage{threeparttable}
\usepackage{amssymb}
\usepackage{graphicx}
\usepackage{wrapfig}
\usepackage{natbib}
\bibpunct{(}{)}{;}{a}{}{,}  

\newcommand{\eqb}{\begin{eqnarray}}
\newcommand{\eqe}{\end{eqnarray}}
\newcommand{\me}{m_{\rm e}}
\newcommand{\mpr}{m_{\rm p}}
\newcommand{\estar}{\epsilon_{\star}}
\newcommand{\ep}{\epsilon_{\rm peak}^{\rm syn}}
\newcommand{\epssc}{\epsilon_{\rm peak}^{\rm ssc,1}}
\newcommand{\epsec}{\epsilon_{\rm peak}^{\rm ssc,2}}
\newcommand{\emax}{\epsilon_{\rm max}}
\newcommand{\emin}{\epsilon_{\rm min}}
\newcommand{\emxf}{\epsilon_{\rm max}^{\rm ssc,1}}
\newcommand{\emnf}{\epsilon_{\rm min}^{\rm ssc,1}}
\newcommand{\sth}{\sigma_{\rm T}}
\newcommand{\Bcr}{B_{\rm cr}}
\newcommand{\nel}{n_{\rm e}}
\newcommand{\ns}{n_{\rm s}}
\newcommand{\nssc}{n_{\rm ssc}^{(1)}}

\newcommand{\us}{u_{\rm s}}
\newcommand{\ub}{u_{\rm b}}
\newcommand{\gc}{\gamma_{\rm c}}
\newcommand{\linj}{\ell_{\rm e}^{\rm inj}}
\newcommand{\lpinj}{\ell_{\rm p}^{\rm inj}}
\newcommand{\lb}{\ell_{\rm B}}
\newcommand{\lsyn}{\mathcal{L}_{\rm peak}^{\rm syn}}
\newcommand{\lsscf}{\mathcal{L}_{\rm peak}^{\rm ssc,1}}
\newcommand{\lsscs}{\mathcal{L}_{\rm peak}^{\rm ssc,2}}

\begin{document}

\title{One-zone SSC model for the core emission of Centaurus A revisited}
\author{Petropoulou, M., Lefa, E., Dimitrakoudis, S. and Mastichiadis, A.}
\institute{Department of Physics, University of Athens, Panepistimiopolis, GR 15783 Zografos, Greece}
\date{Received.../Accepted...}

\abstract{}
{We investigate the role of the second synchrotron self-Compton (SSC) photon generation to 
the multiwavelength emission from the compact regions of sources
that are characterized as misaligned blazars. For this, we focus on the nearest high-energy emitting radio galaxy Centaurus~A and we
revisit the one-zone  SSC model for its core emission.}
{We have calculated analytically the peak luminosities of the first and second SSC components by, first, 
deriving the steady-state electron distribution in the presence of synchrotron and SSC cooling and, then, by
using appropriate expressions for the positions of the spectral peaks. We have also tested our analytical results
against those derived from a numerical code where the full emissivities and cross-sections were
used. 
}
{We show that the one-zone SSC model cannot account for the core emission of Centaurus A  above a few GeV, where
the peak of the second SSC component appears. 
We, thus, propose an alternative explanation for the origin of the high energy ($\gtrsim 0.4$ GeV) and TeV emission,  where these
are attributed to the radiation emitted by a relativistic proton component through photohadronic interactions with the photons produced by the 
primary leptonic component.
We show that the required proton luminosities are not extremely high, e.g. $\sim 10^{43}$~erg/s, 
provided that the injection spectra are modelled by a power-law with a high value of the lower energy cutoff.
Finally, we find
that the contribution of the core emitting region of Cen~A to the observed neutrino and ultra-high
energy cosmic-ray fluxes is negligible.
}
{}

\titlerunning{SSC model for Cen~A revisited}
\authorrunning{M. Petropoulou et al.}
\maketitle

\section{Introduction}
Centaurus A (Cen~A) is the nearest radio galaxy to earth with a luminosity distance $D_{\rm L} \simeq 3.7$ Mpc
\footnote{Although there is still considerable debate on
its distance, see e.g. \citealt{ferrarese07, majaess10, harrisetal10} we will adopt this value as a representative one.}
and, therefore, one of the best labarotories 
for studying the physics of radio lobes, relativistic outflows, shock formation, thermal and non-thermal emission mechanisms.
Due to its proximity, emission from the extended lobes and jet as well as from its
nucleus has been detected across the electromagnetic spectrum (see e.g. \cite{israel98} for a review). 
In radio wavelengths it has an FRI morphology \citep{fanaroffriley74}, while in 
higher energies (X-rays) is regarded as a misaligned BL Lac object \citep{morganti92, chiaberge01} 
in agreement with the unification scheme of Active Galactic Nuclei (AGN) \citep{padovaniurry90, padovaniurry91}. 
Although the angle between the jet axis and our line
of sight is large, it is still not well constrained mainly due to the assumptions used in
its derivation (see e.g. \citealt{hardcastle03}); it ranges between $15^{\circ}$
 \citep{hardcastle03} up to $50^{\circ}-80^{\circ}$ \citep{tingay98}. 
 
Gamma-ray emission ($\sim 0.1-10$ GeV) from Cen~A has been detected by EGRET \citep{hartman99} but the identification of the $\gamma$-ray
source with the core was rather uncertain due to large positional uncertainties.
The recent detection of very high energy (VHE) emission  ($\sim$ TeV) from the core of Cen~A by 
H.E.S.S. \citep{aharonian09} along with the {\sl Fermi} satellite observations above 100 MeV from the core \citep{abdo10a} and 
X-ray data from various telescopes  make
 now possible the construction of a well sampled Spectral Energy Distribution (SED) for its nuclear emission\footnote{We note
that high-energy (HE) emission
was also detected from the radio lobes of Cen~A
\citep{abdo10b}, while a recent analysis by \cite{yang12} shows that this emission extends beyond the radio lobes. However, 
we will not
deal with the lobe emission in the present work.}
which requires physical explanation. Whether the HE/VHE core emission
originates from very compact or extended regions is still unclear because of lacking information regarding the variability in the GeV/TeV energy ranges and 
of 
the current resolution of $\gamma$-ray instruments. This complicates further the attempts of fitting the multiwavelength (MW) core emission.

The one-zone synchrotron self-Compton (SSC) model 
is one of the most popular emission models due to its simplicity
and to the small number of free parameters. In the past it has been successfully
applied to the SEDs of various blazars -- see e.g. \cite{ghisellini98, celotti08} for steady-state 
models and \cite{mastkirk97, boettcherchiang02} for time-dependent ones. Note, however, that 
rapid flaring events and 
recent contamporeneous MW observations of blazars pose problems to homogeneous SSC models \citep{begelman08, boettcher09, costamante10}.
If FRIs are indeed misaligned BL Lac objects, then one expects that the 
one-zone SSC model applies also successfully to their MW emission (see e.g. \cite{abdo09b} for M87 and  \cite{abdo09a} for NGC~1275).
We note, however, that alternative emission models have also been proposed (e.g. \cite{gianniosetal10} for M87).
In the case of Centaurus A it is still the leading interpreting scenario for the core emission,
 at least below the TeV energy range \citep{chiaberge01, abdo10a, roustazadeh11}.

However, 
there is a subtle point that must be taken into account when applying one-zone models to FRIs:
due to the large viewing angle the Doppler factor $\delta$ cannot take large values (in most cases $\delta < 5$) in contrast to blazars
where typical values are $\delta \sim 20-30$, while even higher values ($40 <\delta < 80 $) appear in the literature \citep{konopelko03, aleksic12}.
Thus, unless the observed $\gamma$-ray luminosity of FRIs is by a few orders of magnitude lower than the one of blazars, the 
injection {power} of relativistic  radiating electrons must be high enough to account for it\footnote{We remind that
 for emission from a spherical region that moves with a Doppler factor $\delta$, the relation between
the luminosity as measured in the rest-frame $L'$ and the observed one $L_{\rm obs}$  is $L_{\rm obs}=\delta^4 L'$.}. 
The above imply that in cases where the radius of the emitting source is not very large,
higher order SSC photon generations
may, in general, contribute to the total SED and  are not negligible as  in the case of blazars.

 In the present work we focus on Cen~A as a typical example of a misaligned blazar. We show that
the simple homogeneous SSC model cannot fully account for its MW core emission due to the emergence of
the second SSC photon generation.
We, therefore, present an alternative scenario where the 
SED up to the GeV energy range is attributed to SSC emission of primary electrons, while the GeV-TeV emission itself
 is attributed to photohadronic processes.

The present work is structured as follows: in Section 2 we calculate analytically using certain approximations the peak luminosities of
the synchrotron and SSC (first and second) components for parameters that are relevant to Cen~A;
in the same section we test our results against those obtained from a numerical code that employs the full
expressions for the cross sections and emissivities of all processes. In Section 3 we show
the effects that the second SSC component has on the overall one-zone SSC fit of the MW core emission of Cen~A.
 In Section 4 we introduce a relativistic proton distribution in addition to the primary electron one, and present
the resulting leptohadronic fits
to the emitted MW spectrum; we also discuss 
the resulting neutrino and ultra-high energy cosmic ray emission.
Finally, we conclude in Section 5 with a discussion of our results. 

\section{Analytical arguments}
The calculation of the steady-state electron distribution in the case
of a constant in time power-law injection under
the influence of synchrotron and SSC (in the Thomson regime) cooling
can be found in \cite{LM13} -- hereafter LM13. 
However, in the present section and for reasons of completeness,
we derive the analogous solution for monoenergetic
electron injection. On the one hand, this choice 
significantly simplifies our analytical calculations. On the other hand,
it is justified since the power-law photon spectrum in the range $10^{13}-10^{15}$ Hz 
is very steep and it can be therefore approximated by the synchrotron cutoff emission of a monoenergetic electron
distribution.
\subsection{Steady state solution for the electron distribution}
 We assume that electrons are being injected at  $\gamma_0$ and subsequently cool down 
 due to synchrotron and SSC losses. Here we assume that all 
scatterings between electrons and synchrotron photons occur in the Thomson regime, which is true for parameter values 
related to the spectral fitting of Cen~A (see Sections 2.2 and 3). 
The electron distribution cools down to a characteristic Lorentz factor $\gc$ where
the escape timescale ($t_{\rm e, esc}$) equals the energy loss timescale and it is given by
\eqb
\gc = \frac{3 m_{\rm e} c}{4 \sth t_{\rm e, esc} (u_{\rm B} +u_{\rm s})},
\eqe
where $t_{\rm e, esc}=R/c$,  $R$ is
the size of the emitting region, $u_{\rm B}$ is the magnetic energy density and 
\eqb
\label{ustot}
\us = \me c^2 \int_{\emin}^{\emax} {\rm d}\epsilon \ \epsilon \ns(\epsilon),
\eqe
is the energy density of synchrotron photons. 
The integration limits in eq.~(\ref{ustot}) are $\emax=b\gamma_0^2$ and $\emin=b\gc^2$, 
where $b=B/\Bcr$ and $\Bcr=4.4\times10^{13}$~G. 
In what follows, all energies that appear in the relations will be normalized with respect to $\me c^2$,
unless stated otherwise. Here we assume that $\gc$ is much smaller than $\gamma_0$, which further implies that
the particle escape is less significant  than the energy losses in shaping the electron distribution at
the particular energy range. Thus, 
 the electron distribution $\nel$ at the steady state is described by the kinetic equation below
\eqb
\label{kinetic}
Q(\gamma) = \frac{4}{3 \me c^2}\sth c \frac{\partial}{\partial \gamma}\left[\gamma^2 \nel(\gamma) \left(u_{\rm B} +\us \right)\right]
\eqe
where and
$Q(\gamma)=Q_0 \delta(\gamma-\gamma_0)$ is the injection rate per unit volume of electrons having Lorentz factors in 
the range $(\gamma, \gamma+d\gamma)$.  
Under the $\delta$-function approximation for the synchrotron emissivity, the differential number
density of synchrotron photons is given by -- see e.g. \cite{mastkirk95}, 
\eqb
\label{nsyn}
\ns(\epsilon) =  A_2  \epsilon^{-1/2} \nel\left(\sqrt{\frac{\epsilon}{b}}\right) ,
\eqe
where
\eqb
A_2=\frac{2}{3}R\sth \ub b^{-3/2}.
\eqe
and $\ub$ is the dimensionless magnetic energy density, i.e. 
$\ub=u_{\rm B}/\me c^2$. 
Plugging eqs.~(\ref{ustot}) and (\ref{nsyn}) into eq.~(\ref{kinetic})
we find
\eqb
\label{kinetic2}
Q_0 \delta(\gamma-\gamma_0) = \frac{4}{3} \sth c \frac{\partial}{\partial \gamma}
 \left[\gamma^2 \nel(\gamma) G_{\rm e} \right],
\eqe
where $G_{\rm e}$ depends on the electron distribution as
\eqb
G_{\rm e}=\left(\ub + \frac{4}{3}\sth R \ub \int_{\gc}^{\gamma_0} {\rm d} x \  x^2 \nel(x) \right).
\eqe
 An ansatz for the solution $\nel$ of the 
above integro-differential equation is 
$\nel(\gamma)=k_e \gamma^{-p}$ with $k_e$ and $p$ being the parameters to be determined.
By substituting the above solution into eq.~(\ref{kinetic2}) we find that $p=2$ and
that $k_e$ satisfies the following quadratic equation
\eqb
\left(\frac{4}{3}\sth R \ub \gamma_0 \right) k_e^2 + \ub k_e -\frac{3Q_0}{4\sth c} =0,
\eqe
with $k_e$ being the positive root
\eqb
k_e = \frac{3}{8\sth R\gamma_0} \left(-1+\sqrt{1+\frac{4 Q_0 R \gamma_0}{c \ub}}\right).
\label{keQo}
\eqe
It is more convenient to express $k_e$ in terms of the electron injection compactness $\linj$, which is defined as 
\eqb
\linj = \frac{\sth L_{\rm e}^{\rm inj}}{ 4 \pi R m_{\rm e} c^3,}
\eqe
where $L_{\rm e}^{\rm inj}$
is the total injection luminosity of electrons. 
Using the relation between $Q_0$ and $\linj$ for monoenergetic injection, i.e.  
\eqb
Q_0 = \frac{3\linj c}{\sth R^2 \gamma_0}
\eqe
we find 
\eqb
k_e  =  \frac{3}{8 \sth R \gamma_0} \left(-1+\sqrt{1+\frac{12 \linj}{\lb}}\right),
\label{kele}
\eqe
where the `magnetic compactness' $\lb = \sth R \ub$ was introduced.
There are two limiting cases that can be studied depending on the ratio $12 \linj/\lb$.
\begin{itemize}
 \item Synchrotron dominated cooling or $\linj << \lb/12$ where we find
\eqb
k_e \approx \frac{9\linj}{4 \sth^2 R^2 \ub \gamma_0} + \mathcal{O}\left( (\linj/\lb)^2 \right) 
\eqe
\item Compton dominated cooling or $\linj \gg \lb/12$ where we find 
\eqb
k_e \approx \frac{3}{4} \left(\frac{3 \linj}{ R^3 \sth^3 \ub \gamma_0^2}\right)^{1/2}.
\eqe
\end{itemize}
\subsection{Peak luminosities}
The relation between the electron injection rate and the 
normalization of the distribution at the steady-state (eqs.~(\ref{keQo}) or (\ref{kele})) is
crucial for the correct calculation of the peak luminosities.
The calculation is complete when the proper expressions of the emissivities and of 
the energies where the peaks appear are taken into account. Our results, for each emission component, 
are presented below.

\subsubsection*{Synchrotron component}
 In the optically thin to synchrotron self-asborption regime, which is 
the case considered here,
the differential synchrotron luminosity per unit volume  is given by 
$J_{\rm syn}(\epsilon) = \left(c/R\right) \us(\rm \epsilon)$; we note that the units of 
$J_{\rm syn}$ are ${\rm erg} \ {\rm cm}^{-3} \ {\rm s}^{-1}$  per dimensionless  energy $\epsilon$.
Under the $\delta$-function approximation for the synchrotron emissivity, 
the peak luminosity (per unit volume) of the corresponding component ($\lsyn$)
emerges at $\ep=b\gamma_0^2$ and it is given by
\eqb
\lsyn \equiv \epsilon J_{\rm syn}(\epsilon)|_{\epsilon=\ep} = \frac{2}{3}\sth \me c^3 \ub \gamma_0 k_e 
\eqe
or using eq.~(\ref{kele})
\eqb
\lsyn = \frac{\ub \me c^3}{4 R}\left(-1+\sqrt{1+\frac{12\linj}{\lb}}\right).
\label{peak_syn}
\eqe
We note that if we were to use the full expression for the
synchrotron emissivity (e.g. \citealt{rybicki79}), the peak in a $\nu F_{\nu}$ plot would
appear at a slightly different energy than $b \gamma_0^2$.
\subsubsection*{First SSC component}
For parameter
values  related to the spectral fitting of Cen~A, e.g. for $\gamma_0=10^3$ and $b\sim 10^{-13}$ we find $\gamma_0 \ep = b \gamma_0^3 <<1$, 
 i.e. scatterings between the maximum energy electrons with the whole synchrotron photon distribution occur in the Thomson regime. 
 Under the above assumption the peak luminosity (per unit volume) of the first SSC component ($\lsscf$) emerges at 
\eqb
\epssc = \left\{ \begin{array}{cc}
                 \frac{4}{3} b \gamma_0^4 e^{-\frac{1}{1-\alpha}} &  {\rm for} \ p<3 \\  
                 \phantom{} & \phantom{} \\
                 \frac{4}{3} b \gamma_0^2 \gc^2 e^{-\frac{1}{1-\alpha}} & {\rm for} \ p > 3
                \end{array}
\right.
\eqe
where $\alpha=(p-1)/2$ is the synchrotron spectral index and $p$ is the power-law index
of the electron distribution at the steady state -- see e.g. \cite{gould79}. In our case the energy
of the peak is given by the first branch of the above equation since $p=2$.
The peak luminosity is then given by
\eqb
\label{l1}
\lsscf \equiv \epsilon_1 J_{\rm ssc,1}(\epsilon_1)|_{\epsilon_1 = \epssc} = \frac{c}{4\pi R}  m_{\rm e}c^2\epsilon_1^2\nssc(\epsilon_1),
\eqe
 where $\nssc$ is  the differential number density of SSC photons (1st generation)  that is given  by
\eqb
\nssc(\epsilon_1) = \frac{4 \pi R}{c} \frac{3 \sth c}{4} \int_{\emin}^{\emax} {\rm d}\epsilon \frac{\ns(\epsilon)}{\epsilon} 
I_{\rm e}(\epsilon_1,\epsilon) ,
\label{nssc1}
\eqe                                 
where 
\eqb
I_{\rm e}(\epsilon_1,\epsilon) = \int_{\max[\gc,1/2\sqrt{\epsilon_1/\epsilon}]}
^{\min[\gamma_0,1/2\sqrt{\epsilon_1/\epsilon}]} {\rm d}\gamma
 \frac{\nel(\gamma)}{\gamma^2} F_C(q,\Gamma_e).
\eqe
Here  $F_C(q,\Gamma_e)$ is the Compton kernel 
\eqb
F_C= 2q\ln q +(1+2q)(1-q) +\frac{1}{2}\frac{\left(\Gamma_e q\right)^2}{1+\Gamma_e q}(1-q)
\label{Fc}
\eqe
and
\eqb
\Gamma_e  =  4 \epsilon \gamma \ {\rm and} \ q  =  \frac{\epsilon_1/\gamma}{\Gamma_e(1-\epsilon_1/\gamma)}.
\eqe
In the Thomson limit, which therefore applies in our case, $\Gamma_e << 1$ and $\epsilon_1 /\gamma <<1$; the Compton kernel 
takes then the simplified form
\eqb
F_C \approx \left(2\frac{\epsilon_1}{4\gamma^2\epsilon}\ln\left(\frac{\epsilon_1}{4\gamma^2\epsilon}\right)+ \frac{\epsilon_1}{4\gamma^2\epsilon}
+1-2\left(\frac{\epsilon_1}{4\gamma^2\epsilon}\right)^2\right).
\eqe
Following \cite{bg70} -- henceforth BG70 -- we assume that  the energies of the scattered photons
lie away from the high- and low-energy cutoffs. Since the integrand of $I_{\rm e}$ is a steep function
of $\gamma$, the upper cutoff does not contribute to the integration, and $I_{\rm e}$ is written as
\eqb
I_e & = &\frac{1}{2} k_e \left(\frac{\epsilon_1}{4\epsilon} \right)^{-3/2}\int_{0}^{1} {\rm d} y y^{1/2}  \left(2 y \ln y +y +1 -2y^2\right) = \nonumber \\
& = & 4 k_e \left(\frac{\epsilon_1}{\epsilon} \right)^{-3/2} C_1
\eqe
where $y=\frac{\epsilon_1}{4\gamma^2 \epsilon}$ and $C_1=0.975 \simeq 1$.
The above expression is then inserted in eq.~(\ref{nssc1}) and we find
\eqb
\nssc(\epsilon_1)= 8 \pi R^2 \sth^2 k_e^2 u_b b^{-1/2} C_1 \ln \Sigma_1 \epsilon_1^{-3/2}, 
\label{nssc1_final}
\eqe
for $4 b \gc^4 <\epsilon_1 < 4 b\gamma_0^4$. In the above, $\ln \Sigma_1$ is the Compton logarithm which also depends on $\epsilon_1$. 
In reality, $\ln \Sigma_1$ changes functional form at $\estar = \frac{4}{3}b\gamma_0^2 \gc^2$  but for the case studied here ($p=2$)
the departure of $\nssc$ from a pure power-law with index $-3/2$, at least away from the cutoffs,
is not significant -- see also eqs.(27)-(28) in \cite{gould79}.
Inserting the above expression into eq.~(\ref{l1}) we find
\eqb
\lsscf=\frac{3\sqrt{3}}{16 e}\frac{\ub \me c^3}{R}\left(-1+\sqrt{1+\frac{12\linj}{\lb }}\right)^2
\label{peak1}
\eqe
where we have neglected the factor $C_1 \ln \Sigma_1$.  Whether our choice is justified or not it
will be tested later on, by comparing eq.~(\ref{peak1}) against the results obtained with the numerical code.

\subsubsection*{Second SSC component}
As already mentioned in the introduction, in the case of blazars, higher order scatterings, i.e. between electrons and
SSC photons of the first generation, are negligible (e.g. see \citealt{bloom96}).
On the other hand, SSC modelling of SEDs from radio galaxies requires, in general, high electron compactnesses ($\linj$)  due to
the deamplified radiation; of course, this is a rather qualitative argument since 
the determination of $\linj$ depends also on the absolute value of the
observed flux, the ratio of the peak luminosities of the 
low- and high-energy humps and the size of the emitting region. 
Here we proceed to calculate analytically the peak luminosity of 
the second SSC component, which will be then compared
to the synchrotron and first SSC peak luminosities.

An analogous calculation to that of eq.~(\ref{nssc1}) for the second generation of SSC photons
is, in principle, more complicated because of the Klein-Nishina effects, which for
the parameters considered here, become unavoidable. In fact, the scatterings
of electrons with SSC photons from the first generation occur only partially in the Thomson and Klein-Nishina regimes.
Thus, one must use the full expression of the Compton kernel (e.g. eq.~(2.48) in BG70), which hinders any further analytical calculations.
In order to proceed, however, we used a simplified version of the single electron Compton emissivity
\eqb
j_{\rm ssc,2}(\epsilon_2) = j_0 \delta\left(\epsilon_2 - \frac{4}{3}  \gamma^2 \epsilon_1\right) H\left(\frac{3}{4}-\gamma \epsilon_1\right),
\eqe
where the step-function introduces an abrupt cutoff in order to approximate the Klein-Nishina supression and $j_0 = 4/3 \sth c \gamma^2 u_{\rm ssc,1}$. 
Here $u_{\rm ssc,1} =  \me c^2 \int {\rm d}\epsilon_1 \epsilon_1 \nssc$ and $\nssc$ is approximated by a single 
power-law, i.e. it is given by eq.~(\ref{nssc1_final}) 
without
the logarithmic term.
The differential luminosity of the second SSC component (per unit volume) is then simply
\eqb
J_{\rm ssc,2}  = \frac{4}{3}\sth c \int_{\gc}^{\gamma_{\max}}{\rm d}\gamma  \int_{\emnf}^{\emxf}  {\rm d} \epsilon_1
\ I_1(\epsilon_2,\epsilon_1,\gamma),
\eqe
where 
\eqb
I_1 = \gamma^2 \nel(\gamma) u_{\rm ssc,1}(\epsilon_1)
\delta\left(\epsilon_2 - \frac{4}{3} \gamma^2 \epsilon_1\right) H\left(\frac{3}{4}-\gamma \epsilon_1\right).
\eqe
After making the integration over $\gamma$ we find
\eqb
J_{\rm ssc,2} & = &\frac{\sth c}{\sqrt{3}}u_{\rm ssc,1}^0 k_e \epsilon_2^{-1/2} I_2(\epsilon_2),
\eqe
where 
\eqb
I_2 = \int_{\emnf}^{\emxf} {\rm d}\epsilon_1 \frac{1}{\epsilon_1} 
H\left(\epsilon_2 - 4/3\gc^2 \epsilon_1\right)H\left(E_{\min}-\epsilon_2\right).
\label{jsec}
\eqe
Here $E_{\min}=\min[3/4\epsilon_1,4/3\gamma_0^2\epsilon_1]$, $\emnf = 4/3 b \gc^4$, $\emxf=4/3 b \gamma_0^4$ and
\eqb
u_{\rm ssc,1}^0 = 8 \pi\me c^2 R^2 \sth^2 \ub b^{-1/2} k_e^2.
\eqe
The integral of eq.~(\ref{jsec}) results in the logarithmic term $\ln\Sigma_2$, where $\Sigma_2$ is the ratio of the 
effective upper and lower limits of the first SSC photon distribution, which do not, in principle, coincide with the actual
cutoffs. 
For the purposes of the present study, however, we will neglect the contribution of the logarithmic term. 
In most cases, the scatterings that result in the second SSC photon generation are only partially in the Klein-Nishina regime 
and the 
quantity $\epsilon_2 J_{\rm ssc,2}$ peaks at $\epsec=\gamma_0 e^{-\frac{1}{1-\alpha_1}}$, where $\alpha_1$ is the spectral index of 
the first SSC component and equals to $1/2$ in our work  -- details about the calculation of the SSC peak in different scattering
regimes can be found in LM13.
Thus, the peak luminosity $\lsscs$ is given by
\eqb
 \lsscs & \equiv & \epsilon_2 J_{\rm ssc,2}(\epsilon_2)|_{\epsilon_2=\epsec} \\ \nonumber 
&= & \frac{8\pi}{\sqrt{3} e} \me c^3 R^2 \sth^3 \ub b^{-1/2} \gamma_0^{1/2} k_e^3
\eqe
or after replacing $k_e$
\eqb
\lsscs =\frac{9 \sqrt{3}\me c^3 }{64 e}\frac{\ub}{b^{1/2}R \gamma_0^{5/2}}\left(-1+\sqrt{1+\frac{12 \linj}{\lb}}\right)^3.
\label{peak2}
\eqe
Finally, using eqs.~(\ref{peak1}) and (\ref{peak2}) we define the ratio $\zeta$ as
\eqb
\zeta \equiv \frac{\lsscs}{\lsscf} = \frac{3}{4b^{1/2} \gamma_0^{5/2}}\left(-1+\sqrt{1+\frac{12 \linj}{\lb}} \right).
\label{zeta}
\eqe
In general, if $\zeta>1$ the system can be led to the so-called 
`Compton catastrophe', where 
the peak luminosity of the n$^{\rm th}$--SSC generation is larger than that of the previous one. This succession 
ceases, however, due to Klein-Nishina effects,  as in our case.
If the electron cooling is synchrotron dominated ($\linj << 8.3 \times10^{-2} \lb$), we find $\zeta>1$ if
$\linj > 8.3 \gamma_{0,3}^{5/2} R_{15}^{-1/4} \lb^{5/4}$, where we used the notation $Q_{\rm x}\equiv Q/10^{\rm x}$ in cgs units.
In this regime, both constraints on $\linj$ cannot be satisfied simultaneously for typical values of $\gamma_0$ and $R$, thus, making the Compton catastrophe not relevant.
On the other hand, in the Compton cooling regime ($\linj > 8.3 \times10^{-2} \lb$), $\zeta$ becomes larger than unity
if $\linj > 570 \gamma_{0,3}^5 R_{15}^{-1/2} \lb^{3/2}$. 
\subsection{Tests}
In this paragraph we will compare the analytical expressions given by eqs.~(\ref{peak_syn}),(\ref{peak1}) and (\ref{peak2}) with
those obtained from the numerical code described in \cite{mastkirk95, mastkirk97}, where we 
have used the full expression for the synchrotron and Compton emissivities (c.f. eqs.~(6.33) and (2.48) in \cite{rybicki79} and BG70 respectively).

For the comparison we used
$B=4$~G, $R=10^{17}$~cm, $\gamma_0=10^3$ and three indicative values of 
the electron injection compactness, i.e.  $\linj=10^{-4}, 10^{-3}$ and $10^{-2}$.
Our results are summarized in Table~\ref{table-peaks}, where the first and second value in each row correspond to the numerical and analytical ones
respectively; the ratio $\zeta$ given by eq.~(\ref{zeta}) is also shown. The magnetic compactness for the parameters used here is $\lb = 0.052$.
The first two examples fall into the `synchrotron dominated' regime since $12\linj/\lb = 2.3\times10^{-2}$ and $2.3\times 10^{-1}$ for
$\linj=10^{-4}$ and $10^{-3}$ respectively. Although, for the highest $\linj$ considered here, electrons cool preferentially through the ICS
of synchrotron photons, we still find  $\zeta <1$.

We note that in all cases the differences between our estimates and the numerically derived values are $\sim 2-3$. 
In particular, our approximation for the position of the synchrotron peak (see Section 2.2) is the main cause for the
differences appearing in the first column of Table~\ref{table-peaks}.
In general however, our approximations used for the derivation of eqs.~(\ref{peak_syn}),(\ref{peak1}) and (\ref{peak2}) are reasonable, even in the 
third case of $\linj=10^{-2}$,  where 
$u_{\rm ssc,1}\approx 4 (u_{\rm B}+ u_{\rm s})$;
we remind that our analysis neglects 
the energy density of SSC photons in the electron cooling.

\begin{table}
\centering
\caption{Peak luminosities (in logarithm) of the synchrotron, first and second SSC components along
with the ratio $\zeta$ of the two SSC peak luminosities. In each row the numerical (N) and analytical (A) values are
shown as the first and second values respectively. }
\begin{tabular}{c c c c c c}
\hline 
 $\linj$ & $\log \lsyn$ & $\log \lsscf$ & $\log \lsscs$  &$\zeta$ \\ 
\hline
\hline
$10^{-4}$ & -4.16 {\tiny(N)} & -6.63 & -9.64 &  $9.1 \times 10^{-4}$ \\
\phantom{} & -3.85 {\tiny (A)} &  -6.10 &  -9.20 & \phantom{}  \\
\hline 
$10^{-3}$ & -3.16 & -4.65 & -6.66 &   $8.6 \times 10^{-3}$ \\
\phantom{} & -2.85 &  -4.13 & -6.20 & \phantom{}\\
\hline
$10^{-2}$ & -2.22 & -2.77 & -3.84 & $6.5\times 10^{-2}$ \\
\phantom{} & -1.97 & -2.40 & -3.60 & \phantom{}\\
\hline
\end{tabular}
\label{table-peaks}
 \end{table}

\section{One-zone SSC fit to the core emission of Cen~A}
The emission from the core of Cen~A has the double-peaked shape
observed in many blazars with the low-and high-energy humps peaking at the infrared
and sub-MeV energy ranges respectively \citep{jourdain93, chiaberge01}.
The one-zone SSC model, where relativistic electrons 
are responsible
for the radiation observed in low and high energies has been successfully applied over the years to various blazars 
and recently to FRI galaxies such as M87 \citep{abdo09b}. Although it is also the dominant
interpreting scenario for the core emission of Cen~A it cannot explain the observed SED up to the TeV energy range 
\citep{abdo10a, roustazadeh11}, since 
the emitting region is compact enough for signifant absorption of TeV gamma-rays on the infrared photons produced inside the source \citep{abdo10a, sahakyan13}.
Note also that before the detection of Cen~A at VHE gamma-rays, its whole SED was successfully reproduced by single zone SSC models \citep{chiaberge01}.

In this paragraph we attempt a similar application to the MW emission of Cen A, having in mind though,
that the second SSC photon generation emerges in the SED for (i) 
high enough electron injection compactnesses, (ii) small size of the emitting region and 
(iii) relatively low Lorentz factor of electrons\footnote{Here we imply mononergetic injection
at $\gamma_0$. In the case of steep power-law injection between $\gamma_{\min}$ and $\gamma_{\max}$, the minimum Lorentz factor
of electrons plays the role of $\gamma_0$.} -- see also eqs.~(\ref{peak_syn}), (\ref{peak1}) and (\ref{peak2}).
We note also that the combination of  
the low electron Lorentz factor with weak magnetic fields, as often used in SSC models, implies 
that the second generation Compton scatterings occur only partially in the  Thomson regime. For
this reason, the second SSC component is expected to be much steeper than the first one.

\begin{figure}
\centering
\includegraphics[width=9cm, height=7cm]{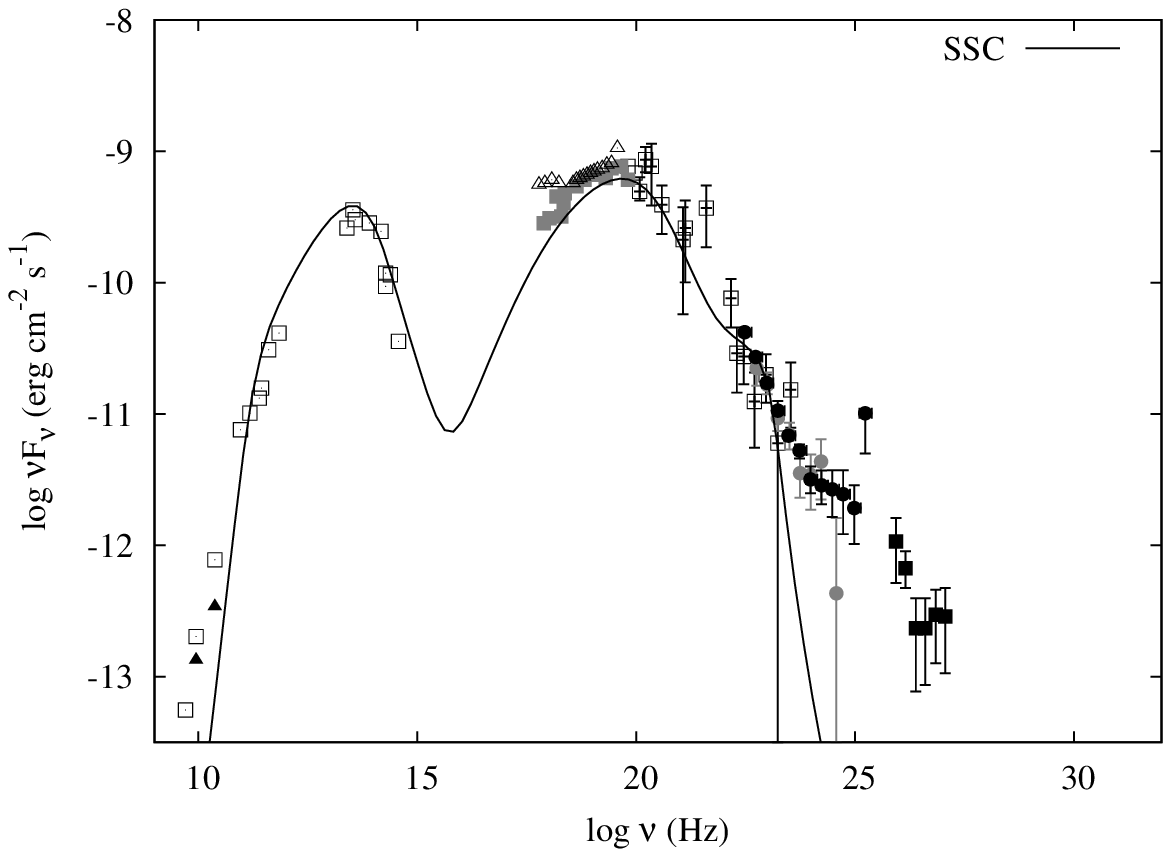}
\caption{SED of the core emission from Cen~A with an one-zone SSC fit. This includes 
non-simultaneous observations from low-to-high frequencies: 
filled triangles (TANAMI VLBI),
grey filled squares ({\sl Suzaku}), open triangles ({\sl Swift}-XRT/{\sl  Swift} -BAT),
grey circles (1-year Fermi-LAT by \citealt{abdo10a}), black circles (4-year Fermi-LAT by \citealt{sahakyan13}), black filled squares (H.E.S.S. by \citealt{aharonian09})
and black open squares are archival data from \cite{marconi00}.
The solid line is our one-zone SSC model fit with same slightly different parameters than those used in \cite{abdo10a}. 
For the parameters used see Table~\ref{table1}.}
\label{cenA}
\end{figure}

Under the assumption of monoenergetic electron injection 
the parameters that must be determined in the context of an one-zone
SSC model are five: $B$, $R$, $\delta$, $\gamma_0$ and $\linj$; 
for power-law and broken power-law injection the unkwnown parameters increase to 
seven and nine respectively -- see e.g. \cite{mastkirk97, aleksic12}.
Because of no 
detections of variability in the HE/VHE regimes, the available observational constraints 
are only four: (i) the ratio of the observed 
peak frequencies $\nu_{\rm peak}^{\rm ssc,1}/\nu_{\rm peak}^{\rm syn}$; 
(ii) the peak synchrotron frequency $\nu_{\rm peak}^{\rm syn} = 3.2\times10^{13}$ Hz;
(iii) the ratio of the observed peak fluxes $\left(\nu F_{\rm \nu}^{\rm syn}\right)_{\rm peak} / \left(\nu F_{\rm \nu}^{\rm ssc,1}\right)_{\rm peak}$;
(iv) the synchrotron peak flux $\left(\nu F_{\rm \nu}^{\rm syn}\right)_{\rm peak} \approx 4 \times 10^{-10}$ erg cm$^{-2}$ s$^{-1}$.
From constraints (i) and (ii) we can determine the injection Lorentz factor of electrons $\gamma_0$ and find a relation between
the  magnetic field strength $B$ and the Doppler factor $\delta$ respectively:
\eqb
\gamma_0 = \sqrt{\frac{3}{4}\frac{\nu_{\rm peak}^{\rm ssc,1}}{\nu_{\rm peak}^{\rm ssc,2}}} = 1.1 \times 10^3
\label{gamma0}
\eqe
and 
\eqb
B = \Bcr \frac{h \nu_{\rm peak}^{\rm syn}}{\delta \gamma_0^2 \me c^2 } = 8 \delta^{-1}\ {\rm G},
\label{Bfield}
\eqe
where we neglected the factor $1+z$ due to the small  value of the redshift ($z=0.00183$). The ratio of the electron to magnetic
compactness is determined by constraint (iii) and eqs.~(\ref{peak_syn}) and (\ref{peak1})
\eqb
\frac{\linj}{\lb} = \frac{1}{12} \left[-1+\left(1+\frac{4e}{3\sqrt{3}}\frac{\left (\nu F_{\rm \nu}^{\rm ssc,1}\right)_{\rm peak} }
{\left(\nu F_{\rm \nu}^{\rm syn}\right)_{\rm peak}} \right)^2 \right] \simeq 5
\label{lelb}
\eqe
Combining constraint (iv) with eqs.~(\ref{peak_syn}), (\ref{Bfield}) and (\ref{lelb}) leads to a relation between $R$ and $\delta$
\eqb
\label{size} 
R  \simeq 10^{15} \delta^{-1} {\rm cm}.
\eqe
Finally, using eqs.~(\ref{lelb}) and (\ref{size}) we find 
\eqb
\linj \simeq 10^{-3} \delta^{-3}.
\eqe
Since the viewing angle of the jet is in the range
$15^{\circ}-80^{\circ}$ the  Doppler factor cannot exceed the value 3.7, whereas
values as low as 0.52 have been used in the literature \citep{roustazadeh11}. From this point on we will adopt the 
representative value $\delta=1$, which
for an angle $30^{\circ}$ implies a bulk Lorentz factor $\Gamma= 7$. 
The derived values ($\gamma_0=1.1\times 10^3$, $B=8$ G, $R=10^{15}$ cm, $\linj=10^{-3}$ and $\delta=1$)
 were then used as a stepping stone for a more detailed fit to the SED, where 
we assumed the injection of a steep power-law electron distribution for better reproducing the
photon spectrum above $10^{13}$ Hz.
The parameter values, which are only slightly different from the analytical estimates, 
are listed in Table~\ref{table1}. 
 In the same table are also listed for comparison reasons the values of the SSC fit by \cite{abdo10a}. We note
that the parameter that differs the most between their fit and ours is the maximum Lorentz factor of the 
electrons. Assuming that the fastest acceleration timescale of electrons is set by their gyration timescale, 
the maximum Lorentz factor
is saturated at $\gamma_{\rm sat} \simeq 5\times 10^7$ due to synchrotron losses in a magnetic field of 6 G. 
It is safe, therefore, to assume that $\gamma_{\rm e, max}=10^6$ (see also \cite{roustazadeh11} for a comment on this point).
Our model SED is shown with solid line
in Fig.~\ref{cenA} and a few features of it are worth commenting:

\begin{table}
\centering
\caption{Parameter values for the one-zone SSC model fit to the SED of Cen~A shown in Fig.~\ref{cenA}.
For comparison reasons, the respective values of the SSC fit by \cite{abdo10a} are also shown.
}
\begin{tabular}{c c c}
\hline 
Parameter & \multicolumn{2}{c}{Model}   \\
\phantom{} & SSC & SSC \citep{abdo10a}\\
\hline
R (cm) & $4\times10^{15}$  & $3\times 10^{15}$\\ 
B (G) & 6 & 6.2\\
$\delta$ & 1 & 1 \\
\hline 
$\gamma_{\rm e, min}$ & $1.3\times10^{3}$  &  300 \\
$\gamma_{\rm br}$ & -- & 800 \\
$\gamma_{\rm e,max}$ & $10^6$\ & $10^8$ \\
\hline
$p_{\rm e,1}$ & -- & 1.8 \\
$p_{\rm e,2}$  & 4.3 & 4.3 \\
\hline
$\linj$ & $6.3 \times10^{-3}$ & $ 8\times 10^{-3}$ \\
$\lb$ & $4.6 \times 10^{-3}$ & $3.7 \times 10^{-3}$\\
\hline
\end{tabular}
\label{table1}
 \end{table}

\begin{enumerate}
\item The steady state electron distribution is completely cooled, i.e.  $t_{\rm syn}(\gamma_{\min})<< R/c$. 
The emission below the peak of the first bump in the SED is attributed to the synchrotron radiation
of cooled electrons below $\gamma_{\min}$ and, therefore, it has spectral index $\alpha=1/2$. 
The inverse Compton scatterings of these low-energy synchrotron photons ($x< b\gamma_{\min}^2$)
with the whole electron distribution occur in the Thomson regime. The resulting
spectrum has also an index $\alpha_1=1/2$ and it explains fairly well the X-ray data from Suzaku and Swift.
\item Although the SSC model  successfully fits the SED from $\sim 10^{10}$ Hz up to $\sim 10^{23}$ Hz, 
it fails in the {\sl Fermi} energy range (grey and black circles in Fig.~\ref{cenA}) due to the 
emergent second SSC photon generation, whose peak appears as a small bump at $\sim 10^{23}$~Hz. In addition,
since most of the scatterings occur in the Klein-Nishina regime, the photon spectrum above that bump steepens abruptly.   
\item The ratio of the second to the first SSC peak luminosities is $\sim 0.05$ as it can be seen from Fig.~\ref{cenA}.
For the parameter values that we derived at the beginning of this section, the analytical expressions given by
 eqs.~(\ref{peak1}) and (\ref{peak2}) predict a ratio $\sim 0.08$, which is in good agreement with the
numerical one.
\item An attempt to fit the SED using the maximum possible Doppler factor ($\delta=3.7$) would
result in smaller values of $R$, $B$ and $\linj$ than those listed in Table~\ref{table1}. This
would suppress electron cooling, i.e.  near/mid-infrared and X-ray observations could not be modelled unless
one would assume the injection of a broken power-law electron distribution.
\end{enumerate}

\section{Addition of a relativistic proton component}
In the previous section we showed that the one-zone SSC model fails to
reproduce the  core emission of Cen~A for energies above a few GeV. A recent analysis
of {\sl Fermi} data from four years of observations resulted in the detection of HE emission
up to $\sim 50$ GeV  \citep{sahakyan13}. It is now believed that this part of the spectrum along with the TeV data
is produced by a second component that originates either from a compact (sub-pc) or from an extended ($\sim$ kpc) region.
Multiple SSC emitting components \citep{lenain08}, non-thermal processes at the black hole magnetosphere \citep{riegeraharonian09}, photopion and photopair
production on background (UV or IR) \citep{kachelriess10} or SSC photons \citep{sahu12}, 
$\gamma$-ray induced cascades in dust torus surrounding the high-energy emitting source \citep{roustazadeh11}, non-thermal emission from
relativistic protons and electrons that are being injected and accelerated at the base of the jet and cool as they propagate along it \citep{reynosoetal10},
are proposed scenarios that fall into  the first category, whereas scenarios such as inverse Compton scattering of background photons
in the kpc-scale jet \citep{hardcastlecroston11} belong to the second one.

Here we propose an alternative explanation for 
 the TeV and the HE emission in the {\sl Fermi} energy range, which may as well be labeled as a `compact origin' scenario.
We assume that inside the compact emission region (e.g. $R=4\times 10^{15}$ cm) relativistic protons, that
have been co-accelerated to high-energies along with the electrons, are being injected in the source.
In a co-acceleration scenario the ratio of the maximum Lorentz factors
achieved by electrons and protons is $\sim \me/ \mpr$, as predicted for example by
first order fermi and stochastic acceleration models (see e.g. \citealt{rieger07}).
For this reason and given that $\gamma_{\rm e, max}=10^6$ we assume that $\gamma_{\rm p, max}=1.8\times 10^9$, which
furthermore does not violate the Hillas criterion since the
corresponding gyroradius is $r_{\rm g} = 4.5 \times 10^{14}$~cm. 
To reduce the number of free parameters in our model we further assume that the accelerated distributions of protons and electrons have the
same power-law index ($p_{\rm p}=p_{\rm e}$), although the resulting photon spectrum is insensitive 
to the exact value $p_{\rm p}$.  

In order to follow the evolution of a system where both relativistic electrons and protons
are being injected with a constant rate in the emitting region we used
the time-dependent numerical code as presented in \cite{dimitrakoudis12} -- hereafter DMPR12. 
The various energy loss mechanisms for the different 
particle species that are included in our code are
\begin{itemize}
 \item Electrons: synchrotron radiation; inverse Compton scattering
\item Protons: synchrotron radiation; photo-pair (Bethe-Heitler pair production) and photo-pion interactions
\item Neutrons: photo-pion interactions; decay into protons
\item Photons: photon-photon absorption; synchrotron self-absorption
\item Neutrinos: no interactions.
\end{itemize}
Photohadronic interactions are modelled using the results of Monte Carlo simulations. In particular, for Bethe-Heitler
pair production the Monte Carlo results by \cite{protheroe96} were used -- see also \cite{mastetal05}. Photo-pion interactions
were incorporated in the time-dependent code by using the results of the Monte Carlo event generator SOPHIA \citep{muecke00}.

\subsection{Photon emission}
\begin{figure}
\centering
\includegraphics[width=9cm, height=7cm]{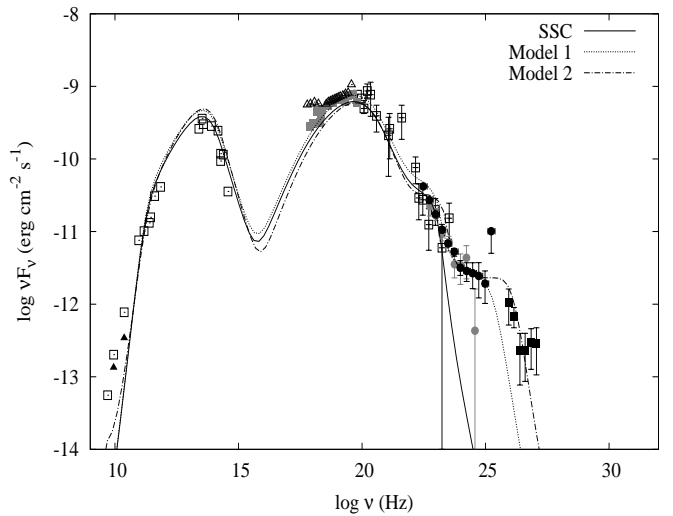}
\caption{Leptohadronic fit of the MW core emission of Cen~A using the parameter sets shown in Table~\ref{table2}. 
Models 1 and 2 are shown with dotted and dashed-dotted lines, respectively. 
For comparison reasons,
the one-zone
SSC fit shown in Fig.~\ref{cenA} is overplotted with solid line. All other symbols are the same as in Fig.~\ref{cenA}.}
\label{cenA_leptohadro}
\end{figure}

\begin{figure}
\centering
\includegraphics[width=9cm, height=7cm]{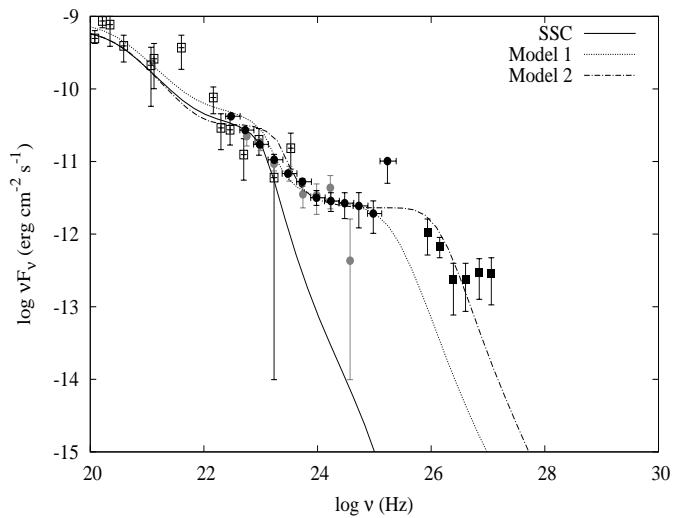}
\caption{Zoom in the $\gamma$-ray energy range of the MW core spectrum of Cen~A. The model spectra are overplotted
with different line types marked on the plot.
}
\label{zoom}
\end{figure}

As a starting template for the parameters describing the primary leptonic component, we first used the one presented
in Table~\ref{table1}. Then, we added five more parameters that describe the relativistic proton component in order to fit
the HE/VHE emission; we refer to this as Model~1. The main difference between Models 1 and 2 is the value of Doppler factor, which is assumed to
be higher in the second model. Subsequently, this affects, as already stated in point (4) of the previous
section, the values of other parameters such as the electron injection luminosity.
The parameters we used for our model SEDs shown in Fig.~\ref{cenA_leptohadro} are listed in Table~\ref{table2}.
In general, the addition of a relativistic proton component 
successfully explains the HE emission 
detected by the {\sl Fermi} satellite by both of our models.  However, the TeV emission detected by H.E.S.S. can be satisfactorily 
explained only by Model~2.  A zoom in the $\gamma$-ray energy
range of the SED along with the model spectra is shown in Fig.~\ref{zoom}.
In what follows, we will first discuss the common features of Models 1 and 2 and, then, we will comment on their
differences.
 
In  both models, gamma-ray emission is attributed to the synchrotron radiation
from secondary electrons produced via Bethe-Heitler pair production and photopion interactions as well as to the  $\pi^0$ decay. 
The hardening of the spectrum at $E\sim 0.4$ GeV, in both cases, 
is caused by photon-photon absorption. This also explains the weak dependence of the resulting model fit on the slope of the proton distribution.

\begin{figure}
 \centering
\includegraphics[width=9cm, height=8cm]{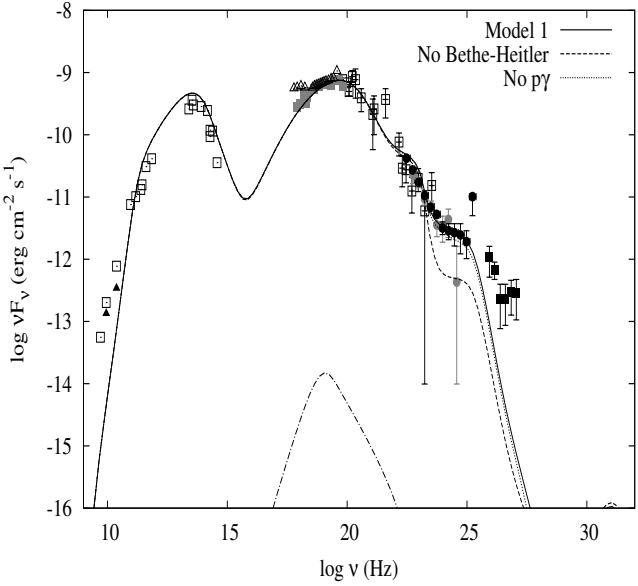}
\caption{Contribution of the photohadronic processes to the high energy part of the spectrum. 
Our model spectra when all processes are included
are shown with solid lines, whereas when photopair and photopion processes are seperately neglected are shown with dashed and dotted lines, respectively.
The dash-dotted curve corresponds to the proton synchrotron emission. For the parameters used see Model~1 in Table~\ref{table2}. }
\label{fig3}
\end{figure}

\begin{table}
\centering
\caption{Parameter values used for our model SED shown in Fig.~\ref{cenA_leptohadro}.}
\begin{threeparttable}
\begin{tabular}{c c c}
\hline 
Parameter & Model 1 & Model 2  \\
\hline
R (cm) & $4\times10^{15}$ & $2.2\times10^{15}$ \\
B (G) & 6 & 3.5\\
$t_{\rm cr}$& $1.3\times 10^5$ s & $7.3 \times 10^4$ s \\
$\delta$ & 1 & 2\\
$\Gamma$ & 7 & 7 \\
$\theta$ & 35$^{\circ}$ & 20$^{\circ}$ \\
\hline 
$t_{\rm e, esc}/t_{\rm cr}$ & 1 & 4 \\  
$\gamma_{\rm e, min}$ & $1.3\times10^{3}$ & $1.3\times10^{3}$\\
$\gamma_{\rm e,max}$ & $10^6$ & $10^6$ \\
$p_{\rm e}$ & 4.3 & 4.5 \\
$\linj$ & $6.3 \times10^{-3}$ & $7.9 \times 10^{-4}$\\
\hline
$t_{\rm p, esc}/t_{\rm cr}$ & 1 & 5 \\
$\gamma_{\rm p, min}$ & $2 \times 10^7$ & $2 \times 10^7$ \\
$\gamma_{\rm p,max}$\tnote{a} & $1.8\times 10^9$  & $1.8\times 10^9$\\
$p_{\rm p}$ & 4.3 & 4.5\\
$\ell_{\rm p}^{\rm inj}$ & $4 \times10^{-6}$ & $7.9\times 10^{-7}$\\
\hline
$u_{\rm r}$ (erg/cm$^3$)\tnote{b} & 12.3 & 2.6\\
$u_{\rm e}$ (erg/cm$^3$) & 1.9 & 2.3 \\
$u_{\rm p}$ (erg/cm$^3$) & 6.8 & 15.4\\
$u_{\rm B}$ (erg/cm$^3$) & 1.4 & 0.5 \\
\hline
$L_{\rm e}^{\rm inj}$(erg/s)\tnote{c}  & $1.2\times10^{43}$ & $1.3\times10^{43}$\\
$L_{\rm p}^{\rm inj}$ (erg/s) & $1.4\times 10^{43}$& $2.4\times 10^{43}$\\
$L_{\rm r}$ (erg/s) & $2.5 \times 10^{43}$ & $2.5\times10^{43}$\\
\hline
\end{tabular}
\begin{tablenotes}
 \item[a] Here $\gamma_{\rm p, \max}\simeq\left(\mpr / m_{\rm e}\right) \gamma_{\rm e,\max}$.
  \item[b] The energy densities refer to the steady state of the system as measured in the comoving frame.
  \item[c] The values refer to observed luminosities.
\end{tablenotes}
 \end{threeparttable}
\label{table2}
\end{table}
In the present treatment we consider only the internally produced photons (synchrotron and SSC)
as targets for photopair and photopion interactions with the relativistic protons, although external
photon fields, such as the radiation from the accretion disk and/or the scattered emission from the Broad Line Region, could also be
important \citep{atoyandermer03}.
The number density of synchrotron photons scales as $n_{\rm syn}(\epsilon) \propto \epsilon^{-3/2}$ for $\epsilon_{\rm cool}^{\rm syn}
 <\epsilon < \ep$, where $\epsilon_{\rm cool}^{\rm syn} \simeq 2.4 \times 10^{-9}$ 
 and  $\ep=2.4\times10^{-7}$. Thus, 
protons with Lorentz factors down to 
$\gamma_{\rm p} \gtrsim 2/\ep \approx 8\times 10^6$ can interact with this photon field through Bethe-Heitler pair production.
Synchrotron photons cannot, however, serve as targets for photopion interactions, since this would require
$\gamma_{\rm p} \ep \gtrsim m_{\pi}/m_{\rm e}$ or equivalently $\gamma_{\rm p}\gtrsim \gamma_{\rm p, max}$. Thus,
pion production is solely attributed to interactions of protons with the SSC photon field  (see also \citealt{sahu12}).
For example, 
protons with Lorentz factors $\gamma_{\rm p} \gtrsim 1.4\times10^3$ and $1.4 \times 10^{7}$ can interact with
the upper ($\emxf \approx 0.2$) and lower ($\emnf \approx 2 \times 10^{-5}$) cutoff of the SSC photon distribution, respectively.
For a fixed proton energy,
 the efficiency of both photopair and photopion interactions depends on the number density of the target field. 
For the particular set of parameters, 
one  expects that interactions between the high-energy part of 
the proton distribution and the low-energy part of the photon
fields are more efficient in the production of $\gamma$-rays.
This is illustrated in Fig.~\ref{fig3}, where the emitted spectra of Model~1 are shown when (i)
 all processes are included (solid line), (ii) Bethe-Heitler pair-production
(dashed line) and (iii) photopion production (dotted line) are omitted. 
It becomes evident that the main contribution to the 
high-energy part of the spectrum comes from the Bethe-Heitler pair production process.
Moreover, the proton synchrotron emission is by many orders of magnitude lower
than the emission from the other components  of hadronic origin -- see dash-dotted line in the 
same figure. 

For the values of $\gamma_{\rm p, min}$ and $p_{\rm p}$ used in the fit, the required injection compactness for 
obtaining an observable high-energy emission signature is $\lpinj=4\times 10^{-6}$ and $7.9\times10^{-7}$ for Models 1 and 2, respectively.
This corresponds to observed injection luminosities $L_{\rm p}^{\rm inj, o} \simeq 1.4  \times 10^{43}$ 
erg/s and $2.4 \times 10^{43}$ erg/s for the two models, respectively\footnote{For the
calculation we used the definition of the proton injection compactness  $\lpinj= L_{\rm p, inj}^{\rm o} \sth / (4 \pi R \delta^4 m_{\rm p} c^3)$, where
the factor $\delta^4$ takes into account Doppler boosting effects for radiation emitted from a spherical volume.}.
For a black hole mass $M_{\rm BH} = 5\times 10^{7} M_{\odot}$ \citep{marconietal06, neumayer10} 
the Eddington luminosity is $L_{\rm Edd} = 6.5 \times 10^{45} \left(M_{\rm BH}/M_{\odot}\right)$ erg/s and, therefore,
the proton injection luminosity in both models is only a fraction of it,
i.e. $ L_{\rm p}^{\rm inj, o} = \xi L_{\rm Edd}$ with $\xi \approx 10^{-3}$.
We note also that the required luminosity of the relativistic proton component is comparable to that of the leptonic one and, therefore, low
compared to the values $10^{47}-10^{48}$~erg/s that are  inferred from typical hadronic modelling of blazars (see e.g. \citealt{boettcher13}).
For the chosen parameters the emitting region is particle dominated with 
$u_{\rm p} + u_{\rm e} \approx \kappa_{\rm i} u_{\rm B}$, where $\kappa_1=6$ and $\kappa_2=36$ for Models 1 and 2, respectively.
 We note also that the radiative efficiency $\eta_{\gamma}$, which we define  as 
 $\eta_{\gamma} = L_{\rm r}/ (L_{\rm e}^{\rm inj} + L_{\rm p}^{\rm inj})$, is high for both models; specifically, the values listed in Table~\ref{table2}
 indicate $\eta_{\gamma,1}=0.98$ and $\eta_{\gamma,2}=0.68$. 

 In both models we have used a high value for the minimum proton Lorentz factor, which 
cannot be explained by any theoretical model of particle injection and acceleration.
However, any effort to extend
such a steep power-law distribution ($p_{\rm p}=4.3-4.5$) down to $\gamma_{\rm p}=1$ is excluded from the energetics.
As an indicative example, we used the parameter values of Model~1 listed in Table~\ref{table2} except for a lower value of the minimum Lorentz factor.
In order to obtain a good fit to the SED for $\gamma_{\rm p, min}=2\times10^{5}$, the required proton injection luminosity increases
by almost three orders of magnitude, i.e. $L_{\rm p}^{\rm inj} = 6\times10^{45}$ erg/s. 
Since there is no physical reason
for such high values of the minimum proton energy, one can interpret it as the break energy
of a broken power-law distribution. In such case, the power-law below the break must be rather flat, e.g. $p_{\rm p} = 1.5-2.0$, in 
order to avoid too high proton luminosities. 
A detailed fit using broken power-law energy spectra lies, however, outside the scope of this work.
In any case, since 
there is no known plausible physical scenario that predicts either high values of $\gamma_{\rm p, min}$
or broken power-law energy spectra with $\Delta p_{\rm p} \ge 2.5$, the 
sub-Eddington proton luminosities listed in Table~\ref{table2} can be considered as a lower limit of those
retrieved using a more realistic proton distribution. 

The key difference between Models 1 and 2 is the assumed value of the Doppler factor. In Model~1, where we did not allow any Doppler boosting of the emitted
radiation ($\delta=1$), we cannot explain the VHE emission. However, by assuming a slightly higher value for the Doppler factor
the intrinsic absorbed spectrum is boosted
by a factor $\sim \delta$ in frequency and of $\sim \delta^3$ in flux, respectively. The boosting effect
when combined with 
the fact that all other parameter values
are of the same order of magnitude as those of Model~1, results in a model spectrum that
satisfactorily goes through the H.E.S.S. data points. 
 In the light of the recent
 analysis of the four-year Fermi-LAT data \citep{sahakyan13} that 
implies a common origin of the HE and VHE emission, 
we believe that Model~2  describes better the emitting region of the core. Note that the connection between
the GeV and TeV emission could not be suggested by the 
previously available one-year Fermi-LAT observations \citep{abdo10a} -- see grey circles in Fig.~\ref{zoom}.

\subsection{Neutrino and UHECR emission}
\begin{figure}
 \centering
 \includegraphics[width=10cm , height=8cm]{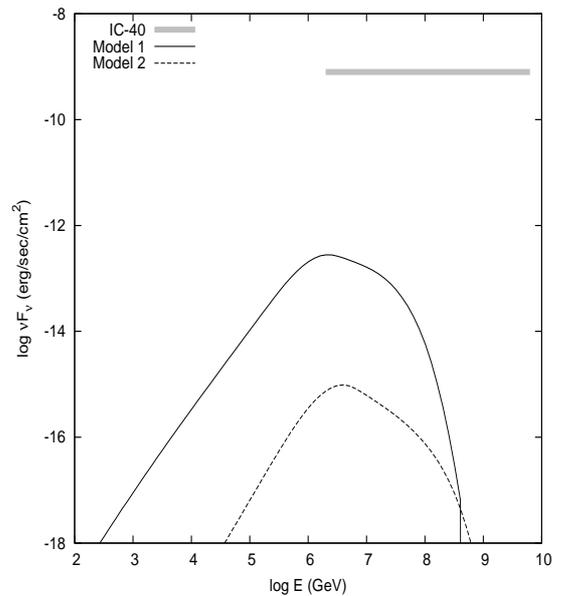}
 \caption{Neutrino spectra of all flavors as obtained in Models 1 (solid line) and 2 (dashed line)
 using the numerical code of DMPR12. The thick solid line shows the IC-40 upper limit.}
 \label{neutrino}
\end{figure}

The detailed neutrino spectra (of all flavors) obtained
using the numerical code of DMPR12 for both models listed in Table~1 are
shown in Fig.~\ref{neutrino}.  The neutrino spectra from both models peak at $\sim 10^6$~GeV, 
 while above that energy they can be approximated as power-laws with  slopes
$p_{\nu} \sim 1.5$ and $\sim 1.6$, respectively.  This is in agreement with
the approximate relation $p_{\nu} \approx (p_{\rm p} - 0.5) / 2.5$ derived in DMPR12. The steepening
of the spectra above $3\times 10^7$ GeV (Model 1) and $10^8$ GeV  (Model 2) 
is due to the cutoff of the proton injection distribution. 
Although photohadronic processes are significant in modelling the photon spectra
above a few GeV, the peak fluxes of neutrinos emitted through 
the charged pion and muon decay are far below the upper limit of the IceCube 40-string (IC-40)
configuration \citep{abbasietal11} -- see grey line in the same figure. 
The neutrino production efficiency that is defined as 
$\eta_{\nu}=L_{\nu}/(L_{\rm e}^{\rm inj}+L_{\rm p}^{\rm inj})$, is approximately $2\times 10^{-5}$ and $2 \times 10^{-7}$ for Models 1 and 2, respectively.
Thus, we find that $\eta_{\nu} << \eta_{\gamma}$, where the radiative efficiency was found to be $\sim 0.8$. 
This differentiates the leptohadronic models presented here from others applied to blazar emission, where neutrino efficiencies as high
as $0.1$ can be obtained (see e.g. \citealt{dimitrakoudis13} for the case of Mrk~421).
In general, there is no case where $\eta_{\nu} \simeq \eta_{\gamma}$ (e.g. \citealt{reimer11}) and such low values 
 are to be expected in cases of strong magnetic fields, weak target photon fields and/or low proton injection compactness; the latter
 applies to our case.
\begin{figure}
 \centering
 \includegraphics[width=10cm , height=8cm]{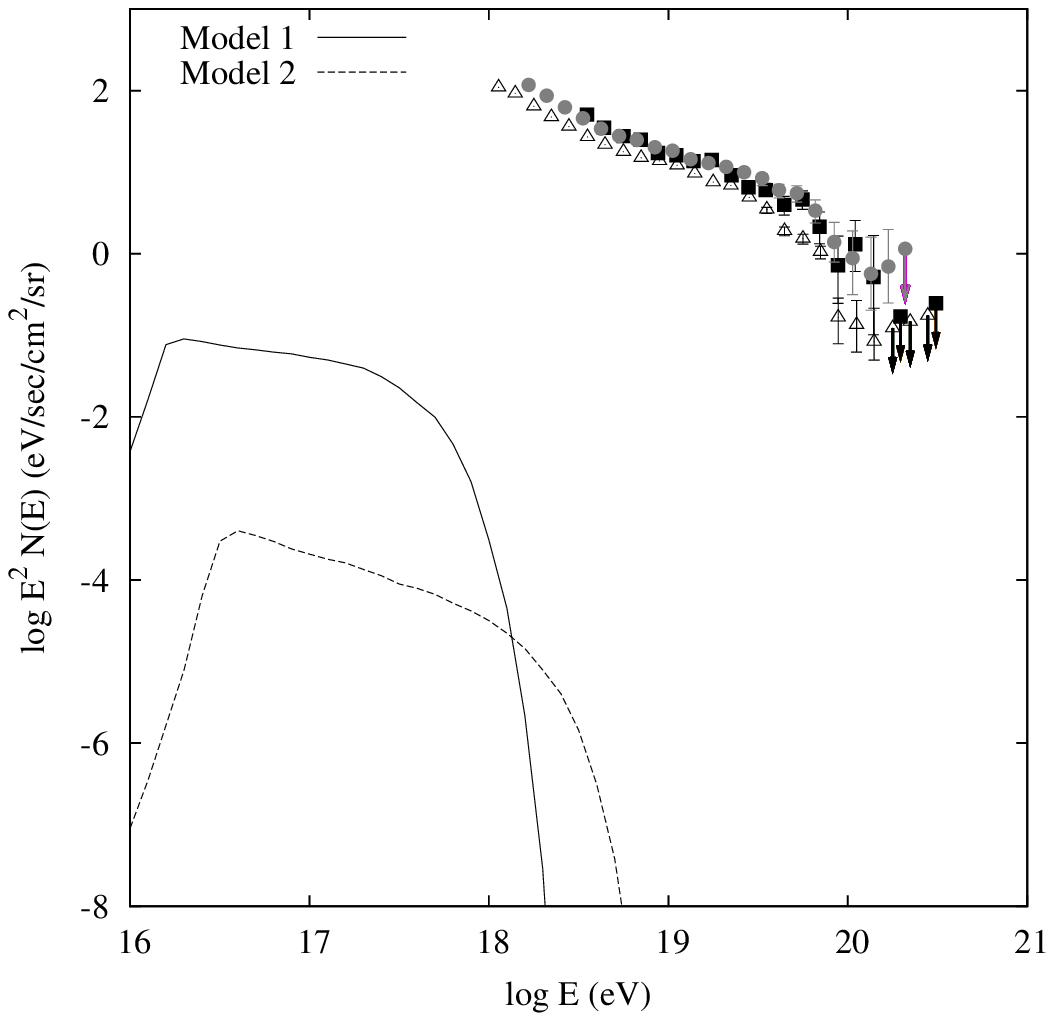}
 \caption{High energy proton spectra resulting from the neutron decay as obtained in Models 1 (solid line) and 2 (dashed line) without taking into
account the effects of diffusion in the intergalactic magnetic field.
The UHECR spectrum as observed by Auger \citep{PA2011}, HiRes-I \citep{HiRes2009} and Telescope Array \citep{TA2013} is overplotted with black
open triangles, grey filled circles and black filled squares, respectively.}
 \label{neutrons}
\end{figure}

Cen~A has been under consideration as a potential source of ultra high energy cosmic rays (UHECR) from as early as 1978 \citep{Cavallo1978}, 
and its proximity to our galaxy compared to all other AGN has even inspired models where it is the sole originator of UHECR \citep{Biermann2012}. 
Recently, the Pierre Auger Observatory (PAO) has shown an excess in UHECR within 18$^\circ$
of Cen~A \citep{PAO2007Sci} and, although that region contains a high density of nearby galaxies, further analysis has 
shown that some of those UHECR may have originated from Cen A itself \citep{Farrar2013, Kim2013}. 
For our two models we
have obtained distributions for both the escaping protons and neutrons. While the former are susceptible to adiabatic energy losses, 
and thus any calculation of their flux would constitute an optimistic upper limit, the latter can escape unimpeded and decay into 
protons well away from the core \citep{KM1989, Begelman1990, Giovanoni1990, AtoyanDermer2003}. In Fig.~\ref{neutrons} 
we have plotted the flux of protons
resulting from the decay of neutrons that escape from the emitting region. 
Since we have not treated cosmic-ray (CR) diffusion in the intergalactic magnetic field, which generally decreases the CR
flux that arrives at earth,
our model spectra should be considered only as an upper limit.
For both models, the peak fluxes are far lower than the observational limit of 
PAO. Although that makes Cen A’s
core an unlikely source of UHECR, those could potentially originate from its lobes instead (e.g. \citealt{Gopal2010}).

\section{Summary/Discussion}

One-zone SSC models for AGN emission have been widely used 
to fit, with varying degrees of success, the SED of blazars.
The discovery of high energy emission from another class of
AGN, i.e. that of radio galaxies, poses new challenges to
these models: if radio galaxies are misaligned blazars, 
then the observed emission should come from a region 
moving with a relatively large angle with respect to our 
line-of-sight. This implies a rather small value for the 
Doppler factor that, for a given flux level of the source,
 can be compensated only by
a large value of the so-called electron compactness parameter.

It is well known that sources with high electron, and consequently
high photon compactnesses, are subject to strong Compton scattering.
This usually leads to higher order generations
of SSC, while, in extreme conditions it might lead 
to the `Compton catastrophe'. As clearly these conditions are not
apparent in the MW spectra of radio galaxies, one could, by reversing the above
arguments, find limits on the parameters used to model the SED of
these sources. 

As an example, in the present paper we have attempted to fit the
SED of the nearby radio galaxy Cen~A, that has been observed both at
GeV and TeV energies. Most researchers agree that the emitting source
is characterized by a low value of the Doppler factor ($\delta\simeq 1-3$).
In order to show the relevance of the first and the second SSC components,
we have calculated analytically in Section 2
the spectral luminosities at the peaks of these
components. Under the assumption that 
all scatterings producing
the first SSC component occur in the Thomson regime, i.e. a condition that
can be easily satisfied in most of the relevant  cases, we found that the SSC 
dominates synchrotron cooling whenever $\linj \ge \lb / 12$, where $\linj$ and 
$\lb$ are the electron and magnetic compactnesses respectively.
The calculation of the luminosity of the second SSC component is more complicated as scatterings
occur in both the Thomson and Klein-Nishina regimes. However, adopting
the oft used cut-off approximation for the latter, we were able to 
find a closed expression for the luminosity which, in addition,
agrees well with numerical calculations -- the same can be said
for the other two components (i.e. synchrotron and first SSC) as
evidenced by Table~\ref{table-peaks}. 

Using the relations described above as a stepping stone, we have 
obtained in Section 3 a fit to the SED of Cen~A. Limiting the Doppler factors
by neccessity to small values, we find that the one-zone SSC model 
can successfully fit the SED up to $10^{23}$ Hz. At that frequency
the peak of the second SSC component appears, which is then followed by a steep power-law segment due to
Klein-Nishina effects.  This causes, typical one-zone SSC modelling to fail at fitting
the high energy observations of Cen~A.

In order to fit the emission at frequencies above $10^{23}$~Hz, we have
introduced, in Section 4.1, a hadronic component which, we assume,
is co-accelerated  to high energies along with the leptonic one. 
Assuming that the two populations share the same characteristics,
i.e. their injection power-laws have the same slope and their
maximum cutoffs are related 
to each other through a simple relation stemming from 
the Fermi acceleration processes, we found that acceptable fits 
to the SED of Cen~A can be  obtained for proton injection luminosities
of the same order of magnitude as the electron one (see Table~\ref{table2}).
Interestingly enough, fits using $\delta=2$ can attribute the TeV
observations to hadronic emission, while fits with $\delta=1$ fail
to do so due to strong photon-photon attenuation.

 In Section 4.1 we have also showed that 
$\gamma_{\rm p, min} \gg 1$ in order to obtain the required radiative efficiency 
of the photohadronic interactions under the assumption of a steep power-law distribution for protons and the 
requirement of a sub-Eddington proton injection luminosity.
On the one hand, such high values of $\gamma_{\rm p, min}$ may be interpreted
as the break energy of a broken-power law at injection. On the other hand, one could, in principle, reconcile the hypothetical low
values of $\gamma_{\rm p, min}$
and the high values of $L_{\rm p}^{\rm inj}$ by considering also as targets for photohadronic interactions external photon fields, such as 
diffuse and/or line emission from the Broad Line Region (BLR). In the case of Cen~A, however, the lack of strong broad emission lines implies
that  these photon fields are negligible \citep{alexander99, chiaberge01}. Another possible photon target field could be
the mid-IR radiation 
that is believed to be associated with cool dust in the nuclear region 
of Cen~A (e.g. \citealt{karovska03}). For the observed fluxes, which range from $1$ to $100$ Jy \citep{israel98, karovska03}, the 
number density of mid-IR photons as measured in the rest frame of the high-energy emitting region is by many orders
of magnitude lower than that of the internally produced synchrotron photons. Thus, incorporating the IR photon
field in the calculations presented here would not lower the requirement of high proton luminosities.

The consideration of relativistic protons in the emitting region is inevitably related
to the neutrino emission, since proton interactions with the 
photon fields present in the source result in charged meson production. In Section 4.2
we have presented the neutrino spectra calculated for both our models.
For the employed parameters the efficiency of pion production
is very low and this can also be seen at the low peak neutrino fluxes
which are by many orders of magnitude below the IceCube upper limit.

Furthermore, high energy neutrons resulting from photopion interactions are an effective
means of facilitating proton escape from the system, as they are unaffected
by its magnetic field and their decay time is long enough to allow
them to escape freely before reverting to protons (e.g. \citealt{KM1989, Begelman1990}). A further
advantage is that they are unaffected by adiabatic energy losses
that the protons may sustain in the system as it expands
\citep{Rachen1998}. Those effects make them excellent candidates 
of UHE protons. For our model parameters, i.e. steep injection proton spectra and small values of the Doppler factor,
the obtained proton distributions peak in the range $10^{16}-10^{17}$~eV, where
the effects of CR diffusion in the intergalactic magnetic field cannot be neglected.
Since in the present work we have not treated CR diffusion, our results should be considered
as an upper limit. Still, these are well below the observed CR flux at such energies.
In the light of recent results suggesting Cen~A  to be the origin of some UHECR events observed 
 by PAO \citep{Farrar2013, Kim2013} and our model results, the core of Cen~A cannot be 
the production site of UHECR.%
%

Our analysis has shown that Cen~A can be explained by means of a leptohadronic 
model as was the case of Mrk~421 \citep{mastetal13}. 
However, contrary to that source,
a one zone SSC model fails to reproduce the SED of Cen~A mainly due to
complications arising from the appearance of the second SSC component.
Although this feature has been overlooked by many researchers it may play
a crucial role  in fitting the SEDs of radio galaxies,
as these require high electron luminosities, making the conditions
very favourable for its appearance.

\section*{Acknowledgements}
We would like to thank the referee Dr. Markus B{\"o}ttcher for his suggestions and
for pointing out several misprints in the manuscript.
This research has been co-financed by the European Union (European Social Fund – ESF) and Greek
national funds through the Operational Program "Education and Lifelong Learning" of the National 
Strategic Reference Framework (NSRF) - 
Research Funding Program: Heracleitus II. 
Investing in knowledge society through the
European Social Fund.  EL acknowledges   financial  support  from  the {\sl thales} project
383549 that  is jointly funded by  the European Union  and the Greek
Government in the framework  of the programme ``Education and lifelong
learning''.
\bibliographystyle{aa}
\bibliography{cenA}
\end{document}